\documentclass[aps,prb,twocolumn]{revtex4-1}
\usepackage{graphicx}
\usepackage{times}
\usepackage{color}
\input pdfcolor.tex
\usepackage{graphicx}
\usepackage{epsfig,color}

\begin{document}
 
\title{Anomalous pseudogap in population imbalanced Fermi superfluids}
\author{Madhuparna Karmakar and Pinaki Majumdar}
\affiliation{Harish-Chandra Research Institute, 
Chhatnag Road, Jhunsi, Allahabad 211 019, India}
 
\begin{abstract}
In a Fermi superfluid increasing population imbalance leads initially to 
reduction of the transition temperature, then the appearance of modulated 
Fulde-Ferrell-Larkin-Ovchinnikov (FFLO) states, and finally the suppression 
of pairing itself. For interaction strength such that the `balanced' system 
has a normal state pseudogap, increasing imbalance reveals anomalous spectral 
behavior. At a fixed weak imbalance (small polarization) the stable homogeneous 
superfluid occurs only above a certain temperature.  The density of states has 
a minimum at the Fermi level, then a weak peak {\it within the gap}, 
and then the large, gap edge, coherence features. On heating, this non monotonic 
energy dependence changes to a more conventional fluctuation driven
pseudogap, with a monotonic energy dependence. At large imbalance 
the ground state is FFLO and `pseudogapped' due to the modulated order. 
It changes to a gapless normal state on heating, and then shows a pseudogap
again at a higher temperature. These weak imbalance and strong imbalance features
both involve effects well beyond mean field theory. We establish them by using 
a Monte Carlo technique on large lattices, motivate the results in terms of the 
pairing field distribution, and compare them to spectroscopic results in the
imbalanced unitary Fermi gas. 
\end{abstract}

\date{\today}
\maketitle

\section{Introduction}

Attractive interaction between fermions lead to pairing and
a superfluid or superconducting ground state. If the interaction
strength is large the pairing effects show up at a high temperature,
$T_{pair}$, via a pseudogap in the density of states, while the
transition to a superfluid occurs at a lower temperature $T_c$.
Between $T_{pair}$ and $T_c$ the pseudogap deepens and for 
$T \lesssim  T_c$ one expects a full gap in the density of states. 
The loss in low energy spectral weight with reducing temperature
is monotonic \cite{randeria_nat2010}.

Since the pairing occurs between time reversed states ${\bf k} \uparrow$
and $-{\bf k} \downarrow$ an imbalance in the population of the
`up' and `down' species, with the total fixed,
 tends to reduce the pairing amplitude
and the $T_c$. The imbalance can be achieved by applying a 
magnetic field, as in the solid state \cite{kumagai}, or by loading 
different numbers of up and down fermions into a trap - as in cold atom
experiments \cite{ketterle2008,hulet,liao}.

The effect of increasing field, $h$, or growing population imbalance,
$P$, is qualitatively similar: it suppresses the $T_c$ of the
homogeneous superfluid (SF), 
promotes a `phase modulated' Fulde-Ferrell (FF) or an
`amplitude modulated' Larkin-Ovchinnikov (LO) state in the intermediate 
$P$ regime, and finally destroys pairing altogether \cite{ff,lo,sarma}.
There are also crucial
differences. For example, while there is a stable ground state
for {\it all applied fields}, there is no stable ground state for
polarization $0 < P < P_{c1}$, 
where $P_{c1}$ is the threshold polarization 
of the FFLO state. At zero temperature the system can either 
be an unpolarised (homogeneous) superfluid (USF) with $P=0$,
or a modulated state with $P_{c1} \le P \le P_{c2}$, or
a homogeneous `normal' state with $P > P_{c2}$. There is
an unstable (phase separation) window $0 < P < P_{c1}$.
This window shrinks
with increasing temperature vanishing at a tricritical
point. The general features are well understood theoretically 
\cite{levin_prl2006,levin2006,torma2007,torma2014}
and have been verified in cold Fermi gases at unitarity 
\cite{ketterle2008, ketterle2007}.

What is less well known is the changing
spectral character of the imbalanced SF with increasing
temperature. We establish this for the two dimensional
attractive Hubbard model at intermediate coupling, where 
the $T_c$ is highest, using a Monte Carlo 
approach. We discover the following:

\begin{enumerate}
\item
At low imbalance the homogeneous superfluid is stable
only above some temperature $T_{un}(P)$ and for $T \gtrsim
T_{un}$ has a pseudogap with a strange 
non monotonic low energy density 
of states. Heating this leads to a crossover to 
a more conventional pseudogap phase with a
monotonic energy dependence.
\item
For $P > P_{c1}$, where the ground
state is FFLO, increasing temperature generates multiple
crossovers: from the `pseudogapped' FFLO to a gapless normal 
state, and, beyond a higher temperature, to another pseudogapped 
normal state. 
\item
The non
paired ground state, with $P > P_{c2}$, is gapless but
on heating develops a weak pseudogap beyond a certain temperature.
This phase is observed in a situation where there is
no order at any temperature, {\it i.e}, $T_c =0$.
\end{enumerate}

\section{Model and method}

We study the attractive two dimensional Hubbard model 
on a square lattice in the presence of a magnetic field:
\begin{equation}
H  = H_0  
- h \sum_{i} \sigma_{iz}  
- \vert U \vert \sum_{i}n_{i\uparrow}n_{i\downarrow} \label{eq1}
\end{equation}
with, 
$ H_0 =  \sum_{ij, \sigma}(t_{ij} - \mu \delta_{ij}) 
c_{i\sigma}^{\dagger}c_{j\sigma}$,
where 
$t_{ij} = -t$ only for nearest neighbor hopping and is zero otherwise.
$ \sigma_{iz} = (1/2)(n_{i \uparrow} - n_{i \downarrow})$.
We will set $t=1$ as the reference energy scale.
$\mu$ is the chemical potential and
$h$ is the applied magnetic field in the ${\hat z}$ direction.
$U > 0$ is the strength of on-site attraction. 
We will use $U/t=4$ and $\mu=-0.2t$ ($n \sim 0.94$).

We have discussed our method of solving the model in
detail elsewhere \cite{tarat,mpk_bp} so we touch on it only briefly.
We use a `single channel' 
Hubbard-Stratonovich (HS) decomposition of the interaction 
in terms of an auxiliary complex
scalar field $\Delta_i(\tau) = \vert \Delta_i(\tau) \vert 
e^{i \theta_i(\tau)}$.
This maps the original interacting  problem to that of non interacting
fermions in a space-time fluctuating field $\Delta_i(\tau)$,
the price to pay is the additional `averaging' over all configurations
of $\Delta_i(\tau)$. Quantum Monte Carlo performs this averaging
without further approximation, 
mean field theory (MFT) restricts $\Delta_i(\tau)$ to a time
independent spatially periodic function, and dynamical mean field
theory (DMFT) treats $\Delta_i(\tau)$ 
as a `single site' time dependent function
$\Delta(\tau)$.

We drop the `time' dependence of $\Delta$, but retain the full
spatial dependence: $\Delta_i(\tau) \rightarrow \Delta_i$. This
corresponds to dropping quantum fluctuations but retaining all
the classical thermal fluctuations. Equivalently one can think
of Fourier transforming the $\Delta_i(\tau)$ into bosonic
Matsubara modes $\Delta_i(\Omega_n)$ and our approach 
corresponds to retaining the $\Omega_n=0$ mode.
This has been called a `static path approximation' (SPA) 
\cite{evenson1970, hubbard}
to the functional integral for the partition function.

SPA retains classical fluctuations of arbitrary magnitude,
 around the saddle point, 
but no quantum ($\Omega_{n} \neq 0$) fluctuation. At $T=0$, since 
the classical fluctuations die off, SPA reduces to standard 
Bogoliubov-de Gennes (BdG) 
mean field theory (MFT). For $T \neq 0$, however,
it considers not only the 
saddle point configuration but {\it all} configurations involving 
classical amplitude and phase fluctuations of the order parameter.  
The BdG equations are solved in all these configurations to 
compute the thermally averaged properties. This approach suppresses 
the order parameter much quicker than MFT. Also, since the 
$\Omega_{n} = 0$ mode dominates the exact partition function, SPA 
becomes exact as $T \rightarrow \infty$.
The approximation leads to the coupled equations:
\begin{eqnarray}
H_{eff}~&=&~H_0 - h \sum_{i} \sigma_{iz} 
+ \sum_{i}(\Delta_{i}c_{i\uparrow}^{\dagger}c_{i\downarrow}^{\dagger}
+ h.c) + H_{cl} \cr
~~&& \cr
P\{{\Delta_{i}}\} &~\propto~& Tr_{c, c^{\dagger}}
e^{-\beta H_{eff}} 
\end{eqnarray}
where $H_{cl} = 
 \sum_{i}\frac{\mid \Delta_{i} \mid^{2}}{U} $
is the stiffness cost associated with the now classical
auxiliary field. The upper equation describes fermions in a 
pairing background $\Delta_i$, while the lower equation
defines the probability distribution  
$ P\{{\Delta_{i}}\} $ associated
with a $\{ \Delta_i \}$ configuration.

We generate samples of $\{ \Delta_i \}$ using a Metropolis
algorithm, diagonalising $H_{eff}$ for the update cost.
In order to make the study numerically less expensive 
the Monte Carlo is implemented using
a cluster approximation \cite{tca}.
After equilibriation at a given $h,~T$ we calculate the
following:
\begin{eqnarray} 
P(h,T) & = & \langle (1/N) \sum_i \langle n_{i \uparrow}
- n_{i \downarrow} \rangle \rangle  \cr
S_{\bf q}(h,T) & = & \frac{1}{N^{2}}\sum_{i, j}
\langle \Delta_{i} \Delta_{j}^{*}\rangle
 e^{i{\bf q}.({\bf r}_{i} - {\bf r}_{j})} \cr
N_{\uparrow}(\omega) & = &
\langle (1/N) \sum_{i,n}\vert u_{n}^i \vert^{2}
\delta(\omega - E_{n}) \rangle 
\nonumber
\end{eqnarray}
where $P(h,T)$ is the  polarization (not be confused with the probability
distribution of the pairing field discussed later),
$S_{\bf q}(h,T)$ is the pairing structure factor, and
$N_{\uparrow}(\omega)$ is   the up spin density of states (DOS).
The down spin DOS is symmetrically shifted.  
$u_n^i$ are the usual BdG
eigenfunctions and $E_{n}$ are the BdG eigenvalues 
in an equilibrium configuration.
Angular brackets indicate thermal average.
Notice that the DOS is also the ${\bf k}$-summed 
spectral function $A_{\sigma}({\bf k}, \omega)$. Formally,
$$
A_{\sigma}({\bf k}, \omega)  =  -(1/\pi)Im G_{\sigma}({\bf k}, \omega)
$$
with, $G_{\sigma}({\bf k}, \omega)~=~lim_{\eta \rightarrow 0}~ 
G_{\sigma}({\bf k}, i\omega_n)\vert_{i\omega_n \rightarrow \omega + i \eta}$
where $G_{\sigma}({\bf k}, i\omega_n)$ is the imaginary frequency 
transform of  
$\langle c_{{\bf k}\sigma}(\tau)c_{{\bf k}\sigma}^{\dagger}(0)\rangle$.
The BdG quasiparticles are related to the original fermions via
the eigenfunction matrix so the Green's function and spectral
functions can be readily computed \cite{degennes}.

\begin{figure}[t]
\centerline{
\includegraphics[width=8cm,height=7cm]{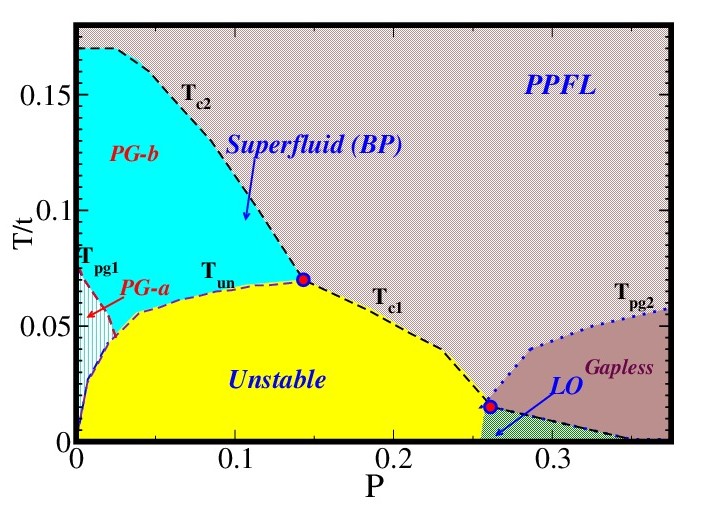}
}
\caption{The polarization $(P)$ - temperature $(T)$ phase
diagram of the intermediate coupling attractive Hubbard
model $(U=4t)$ and $n \sim 0.94$ in two dimensions. The
thermodynamic phases include a homogeneous superfluid, FFLO states,
a partially polarized Fermi liquid (PPFL), and an `unstable' phase
separated region.  The spin resolved density of states can either
have a hard gap, pseudogaps of different character, PG-a or PG-b
(see text), or be gapless.  The superfluid to PPFL transition is
indicated as $T_{c2}$ and $T_{c1}$, for 2nd and 1st order transitions
respectively, and the unstable to superfluid boundary as $T_{un}$.
The PG-a to PG-b crossover within the superfluid is denoted as
$T_{pg1}$ and the gapless to pseudogap crossover at large $P$
denoted as $T_{pg2}$.  The variation of $T$ at fixed $P$ can take
the system through different pseudogap regimes.
}
\end{figure}

We wish to probe results at fixed $P$, for varying $T$, as
would be the case in population imbalanced cold Fermi gases.
Within the Hamiltonian formulation $P$ is a derived quantity,
dependent on $h,T$ and the other model parameters. We therefore
solve the $h-T$ problem first and then {\it construct} a $P-T$
phase diagram out of it.

Since mean field theory is widely used to study the imbalanced 
Fermi gas it is useful to point out that, apart from the gross
overestimate of $T_c$, mean field theory predicts that the
homogeneous SF is {\it always gapped}, and the normal state
{\it always gapless}. The notion of a pseudogap does not
figure in it \cite{randeria_nat2010}.

\section{Results}

\subsection{Phase diagram}

\begin{figure*}[t]
\centerline{
~~
\includegraphics[width=5.6cm,height=5.4cm]{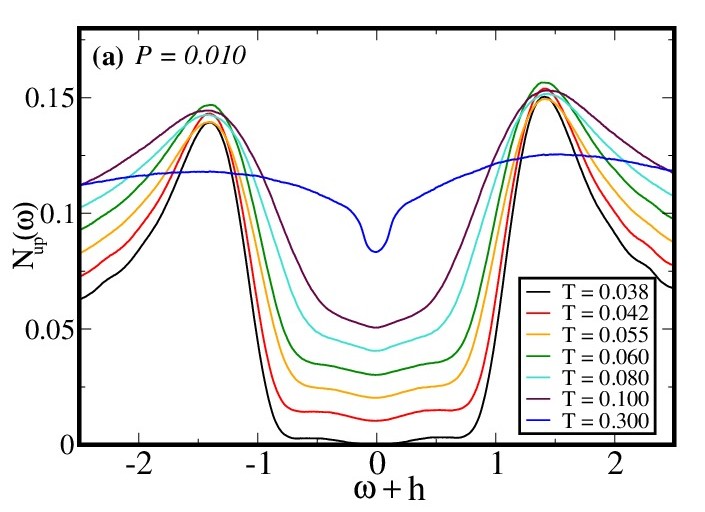}
\hspace{-0.27cm}
\includegraphics[width=5.6cm,height=5.4cm]{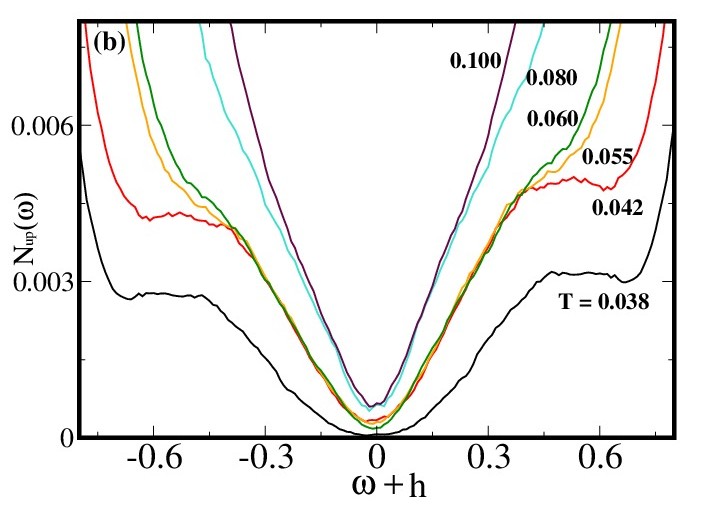}
\hspace{-0.20cm}
\includegraphics[width=5.9cm,height=5.4cm]{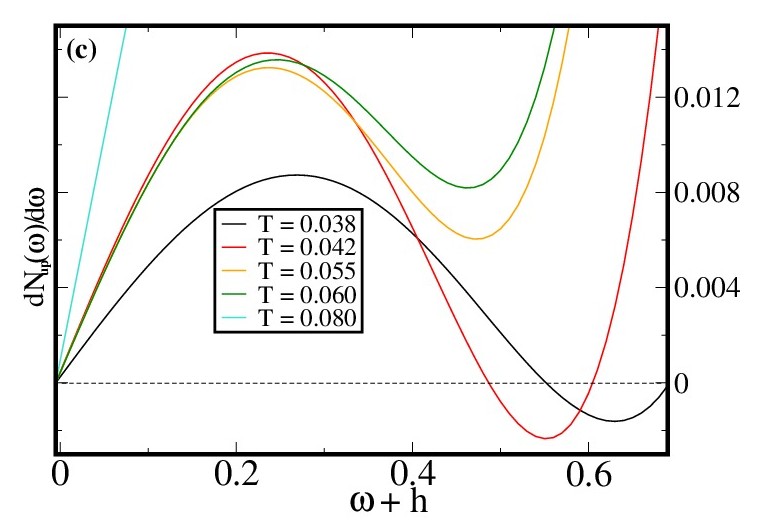}
}
\vspace{-0.1cm}
\centerline{
\includegraphics[width=5.6cm,height=5.4cm]{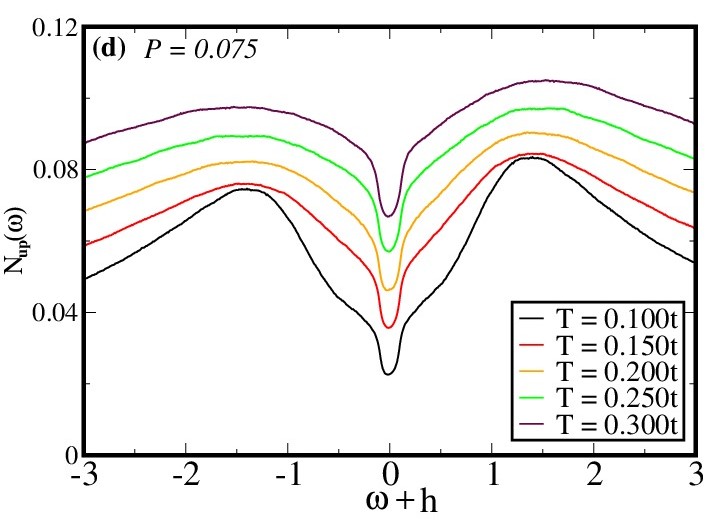}
\hspace{-0.27cm}
\includegraphics[width=5.6cm,height=5.4cm]{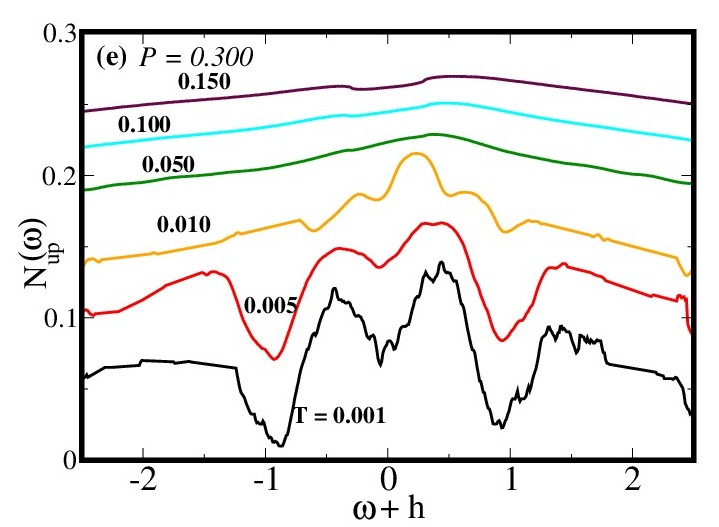}
\hspace{-0.27cm}
\includegraphics[width=5.6cm,height=5.4cm]{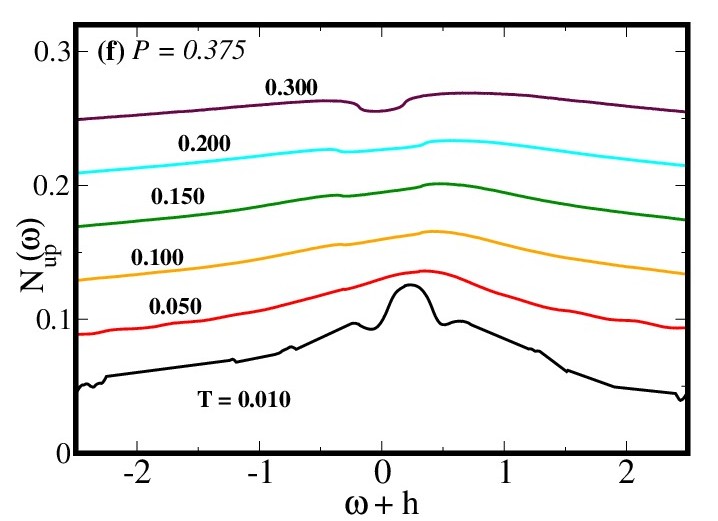}
}
\caption{Temperature dependence of the up spin DOS at (a)~$P
\sim 0.010$, (d)~$P \sim 0.075$, (e)~$P \sim 0.300$ and (f)~$P
\sim 0.375$.  The plots are vertically shifted for clarity and
the $\omega$ axis is shifted by $+h$ to center the gap feature
on the origin.  In (a) the DOS starts with  PG-a character (see
text) at the lowest accessible temperature, transforms to PG-b
on heating (all within the SF phase), then to a pseudogapped
normal state.  In (d) the low $T$ pseudogap weakens but persists
across the SF to PPFL transition and on to high $T$. In (e) the
ground state is `pseudogapped' due to FFLO order. This is lost
on transition to the PPFL but a weak PG reappears at high
$T$. In (f) the PPFL ground state is gapless but develops a
PG on heating. Panel (b) shows the DOS behavior at $P \sim 0.010$,
focusing on the low energy window. Notice the change in character
with increasing $T$. Panel (c) shows the energy derivative of the
DOS at $P=0.01$, the multiple changes  in sign at low $T$ indicate
the multiple minima-maximum structure. The sign changes occur only
for the two lowest $T$, which we have called the PG-a phase. }
\end{figure*}

Fig.1 shows the $P-T$ phase diagram inferred from the $h-T$ phase
diagram (shown later, and also in ref. \cite{mpk_bp}). 
While the thermodynamic phase boundaries are determined based on
$S_{\bf q}(T)$, the various crossovers 
are determined based on the behavior
of the DOS. A suppression in the density of states around the
Fermi level (but not a hard gap) is characterized as a 
pseudogap (PG).
In our case we classify the pseudogaps as PG-a and PG-b depending
on the detailed spectral behavior in the subgap region. 
Considering the $\omega \ge 0$ window, 
PG-a has a minimum at $\omega=0$, a local maximum at
a low energy $\omega_{max}$, a second minimum at an energy
$\omega_{min}$, say, and then the gap edge coherence features.
PG-b involves a minimum only at $\omega=0$.
We first recapitulate the ground state then 
move to thermal properties.

\subsubsection{Ground state:}

In terms of an applied magnetic field, 
the ground state is a homogeneous unpolarised SF
for fields $h < h_{c1} \sim 0.9t$. At $h_{c1}$ it makes a 
transition to a finite polarization FFLO 
state, with $P = P_{c1}$.
The FFLO regime is from $h_{c1}$ to $h_{c2}$, with the polarization
growing from $P_{c1}$ to $P_{c2}$, and beyond $h_{c2}$ the
ground state has no pairing and is a partially polarized Fermi
liquid (PPFL). 

In terms of polarization, at $T=0$ the entire $0 <  h <  h_{c1}$ region,
with $P=0$, collapses to the origin. The polarization regime
$0 < P < P_{c1}$ is unstable, with no homogeneous phase
allowed, followed by the FFLO and PPFL at higher $P$. The $P=0$ 
superfluid is gapped, the FFLO has `pseudogaps', {\it i.e},
depression in its DOS due to the spatial modulation, while the 
PPFL is gapless.

\subsubsection{Thermal properties:}

At finite $T$ the homogeneous SF does allow finite polarization
and is called the `breached pair' (BP) state \cite{mpk_bp}. 
It occupies an increasing $P$ window, as $T$ increases, at the expense of
the unstable region. The temperature above which the SF is stable
is $T_{un}(P)$. At an even higher temperature, $T_{c2}$, superfluidity 
is lost through a Berezinskii-Kosterlitz-Thouless (BKT) transition. 
We infer the $T_c$ scale from $S_0(h,T)$.
 
The upper right edge 
of the unstable region defines the scale, $T_{c1}$,
for a first order
transition between the SF and the normal state. 
The FFLO has a very low $T_c$ and transits to a gapless normal
state on heating. This gapless phase defines a large part of
the high $P$ low temperature window.

The SF phase can be gapped or pseudogapped as Fig.1 shows, the
FFLO is pseudogapped, while the PPFL is gapless at low $T$ and
pseudogapped at high $T$. We will discuss the origin of these 
behavior in the discussion section, but the phase
diagram indicates that varying temperature at a fixed polarization
can lead to multiple changes 
in spectral character. This is directly
relevant for cold Fermi gases where one works at fixed imbalance
rather than a fixed applied field \cite{ketterle2008, hulet, liao}.

\subsection{Density of states}

Fig.2 shows the spin resolved DOS for fixed $P$  
cross sections through the $P-T$ phase diagram in Fig.1.
Two of the panel, (a) and (d), traverse the BP part of the
phase diagram, (e) starts with a FFLO ground state, while
(f) has a PPFL ground state. The ground states vary widely,
we discuss the cases  one by one.

Fig.2(a) is for $P = 0.01$.
The $T=0$ state in this case would be in the unstable window
and stable phases are defined in this case only for $T > 0.035t$.
All the way from this temperature to $T_c \sim 0.17t$ the 
system is a BP superfluid. The behavior of the DOS, however,
changes multiple times. For $0.035t < T < 0.08t$ the (shifted)
DOS has its absolute minimum at $\omega=0$ but then a local
maximum at a small scale $\omega_{max}$ (see plot) and a 
local minimum after that at $\omega_{min}$ and then the sharp
rise at the gap edge. While subgap density of states by itself
is not surprising this additional feature 
is unusual.
Increasing $T$ beyond $\sim 0.08t$ leads to a more common PG
with a monotonic increase in $N(\omega)$ as $\omega$ increases.

Fig.2(b) focuses on the low energy behavior of the data in 2(a).
The lower scales clearly reveal the non monotonic DOS at low $T$.
We fitted this behavior to an approximate form $N(\omega)
= a + b\omega^2 + c \omega^4 + d \omega^6$, and, after
extracting the coefficients $a,~b,~c,~d$, plotted the derivative
$dN/d\omega$ in Fig.2(c). The multiple zero crossings in the
low $T$ data highlight the multiple extrema.

\begin{figure}[t]
\centerline{
\includegraphics[width=8.0cm,height=7.0cm]{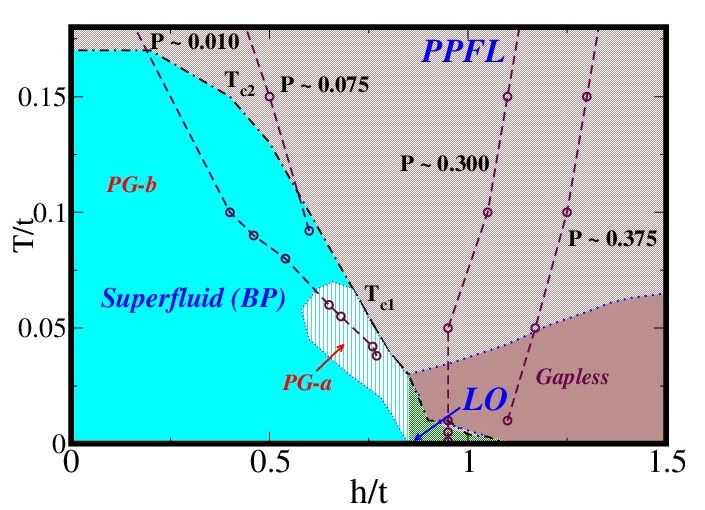}
}
\caption{Magnetic field-temperature $(h-T)$ phase diagram showing
the varying $h(T)$ that is needed to maintain the polarization
fixed at the values chosen in Fig.2. In general when pairing effects
are strong and the low $T$ phase is a BP superfluid the magnetic
field has to sharply rise with reducing $T$ to maintain $P$ fixed.
This is clearly visible for $P=0.01$ and weakly so for $P=0.075$.
In the higher field regime, where the low $T$ state is FFLO or PPFL
the magnetic field needs to increase somewhat with {\it increasing}
$T$, as in a normal metal, to maintain a fixed $P$.  The shaded
region within the BP phase roughly correspond to the anomalous PG
phase (see text for details).  }
\end{figure}

The result in Fig.2(d) is at $P=0.075$.
The stable BP window
at this $P$ starts at $T \sim 0.07t$. Heating this BP state leads to a 
2nd order transition at $T_c \sim 0.14t$. All the way from
$T \sim 0.07t$ through $T_c$ to high temperature in the PPFL phase
the system is pseudogapped, with the character that we called
PG-b, with monotonic $N(\omega)$. 
The coherence peaks vanish, the depth of the pseudogap lessens,
and the low energy weight increases with increasing $T$. 

Fig.2(e) is for $P \sim 0.30$.
The ground state in this case is an axial stripe LO phase,
with ${\bf Q} = (\pi/3,0)$. 
The periodic modulation of $\Delta_i$ 
in the low $T$ state leads to multiple bands and a 
redistribution of the tight binding spectral weight.
The presence of a finite ${\bf Q}$ modulation leads to 
pairing of fermions between $\vert {\bf k}{\uparrow}\rangle$
and $\vert -{\bf k} \pm {\bf Q} {\downarrow}\rangle$ states 
(rather than the 
time reversed $\vert {\bf k} {\uparrow}\rangle$ and 
$\vert  {\bf -k} {\downarrow} \rangle$) and give rise to 
multiple branches in the dispersion. Some of these 
branches cross the Fermi level leading to a finite 
but depressed DOS around $\omega \sim 0$.
This is not a fluctuation induced pseudogap but a
`band structure' effect.
The $T$ window immediately above $T_c$, $T \sim 
[0.01t,~0.05t]$, in
Fig.2(e), is gapless. The pairing field is small and
disordered and does not affect the DOS noticeably. However,
at the two highest temperatures in Fig.2.(e) a weak pseudogap
again emerges. This is due to the thermally induced growth
in the mean magnitude $\langle \vert \Delta_i \vert \rangle$
with the phase variables $\theta_i$ remaining random.
At this polarization one has a pseudogap-gapless-pseudogap
re-entrance with increasing temperature.

\begin{figure*}[t]
\centerline{
\includegraphics[width=7.0cm,height=6.0cm]{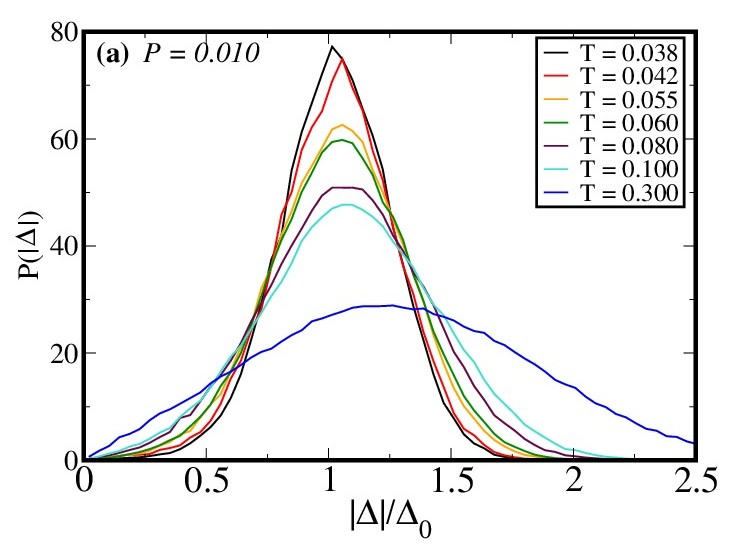}
\hspace{-0.27cm}
\includegraphics[width=7.0cm,height=6.0cm]{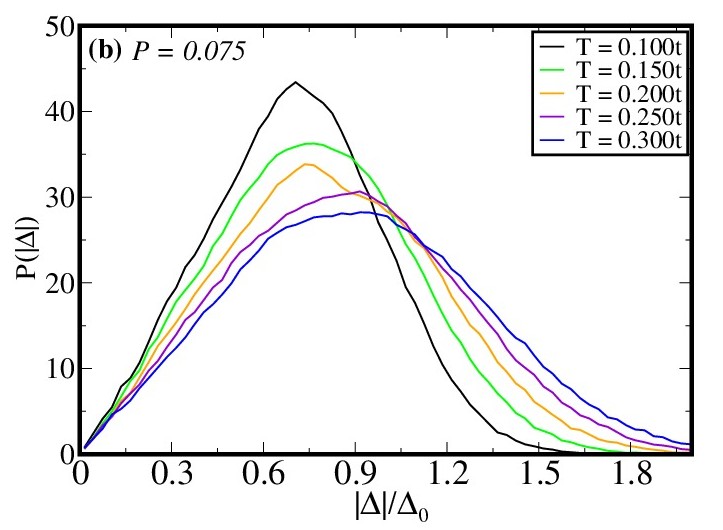}
}
\vspace{-0.1cm}
\centerline{
\includegraphics[width=7.0cm,height=6.0cm]{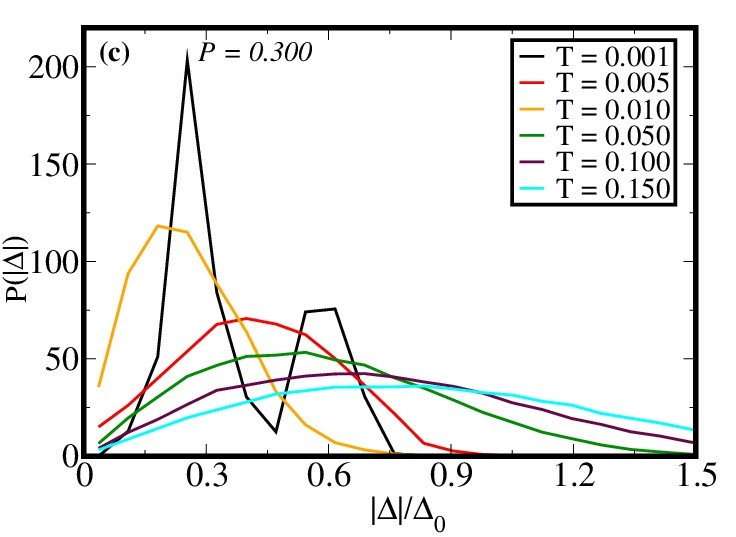}
\hspace{-0.27cm}
\includegraphics[width=7.0cm,height=6.0cm]{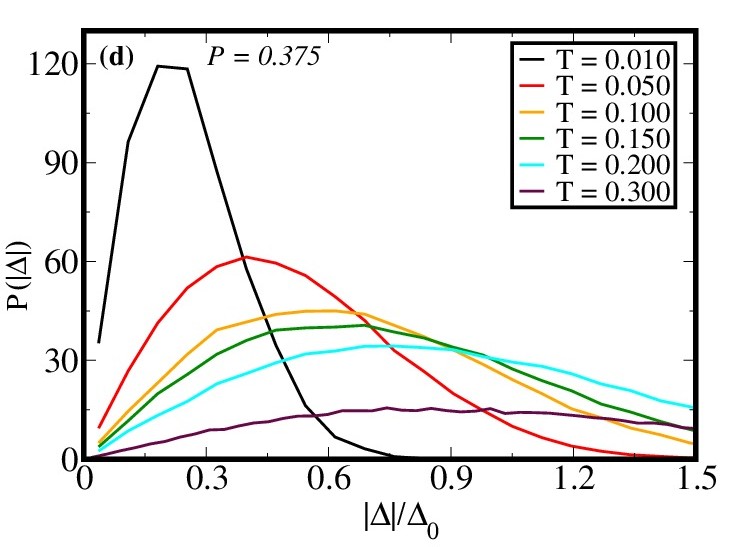}
}
\caption{The $T$ dependence of the
pairing field distribution for different $P$.
The x axis is normalized with respect
to the $T=0$ unpolarised SF amplitude. The panels are for (
a)~$P~\sim 0.01$, (b)~$P \sim 0.075$, (c)~$P \sim 0.300$ and (d)~$P
\sim 0.375$.}
\end{figure*}

Finally, Fig.2(f) at $P=0.375$ has a non ordered ground state
with no $\Delta_i$ at any site. This is a system at strong
interaction, $U=4t$, where the imbalance suppresses pairing
at $T=0$. As a result the low temperature phase is gapless.
With increasing $T$ the $\langle \vert \Delta_i \vert \rangle$
grows quickly, as in the FFLO window, and leads to a pseudogap
for $T > 0.07t$. 
This is a simple instance of a pseudogap emerging without the
presence of any order in the low temperature state.

\section{Discussion}

We first discuss the origin of the effects seen in Fig.1 and
Fig.2, and then earlier theory efforts and experimental data
on this topic.

\subsection{$h-T$ trajectory}

Fig.3 shows the field-temperature phase diagram from which
the $P-T$
phase diagram is derived. The basic phases have 
been discussed earlier in
ref. \cite{mpk_bp}. It is included here mainly to show the
rather unusual $h-T$ paths that lie behind the constant
$P$ cross sections in Fig.1.

At $P=0.01$ the low $T$ part of the trajectory passes close to the
BP-LO phase boundary.  We explored the $h-T$ neighborhood and
discovered that in the window shaded grey the DOS indeed displays
the peculiar features observed in Fig.2(a). The proximity to the
BP-LO phase boundary, at finite $T$, suggests that while much of
the $\Delta_i$ would have (large) values appropriate to the BP
phase, there could be a small fraction of sites that have the
lower values appropriate to the LO phase. We will examine this aspect
in the distributions later.

\begin{figure*}
\centerline{
\includegraphics[width=5.5cm,height=5.0cm]{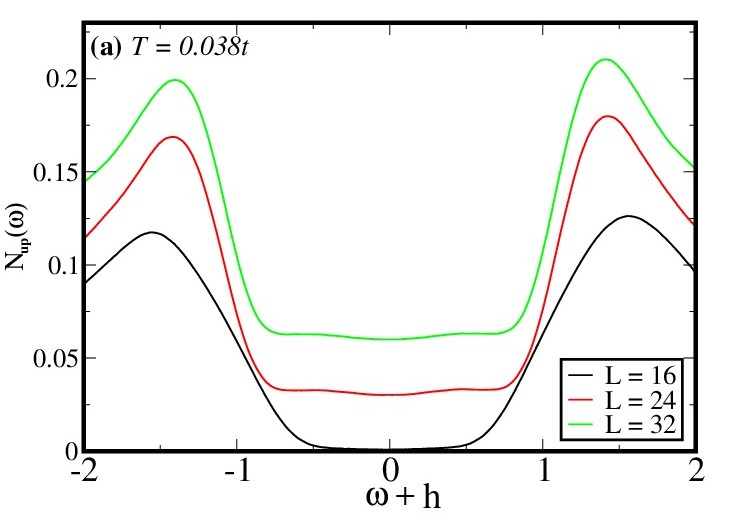}
\hspace{-0.27cm}
\includegraphics[width=5.5cm,height=5.0cm]{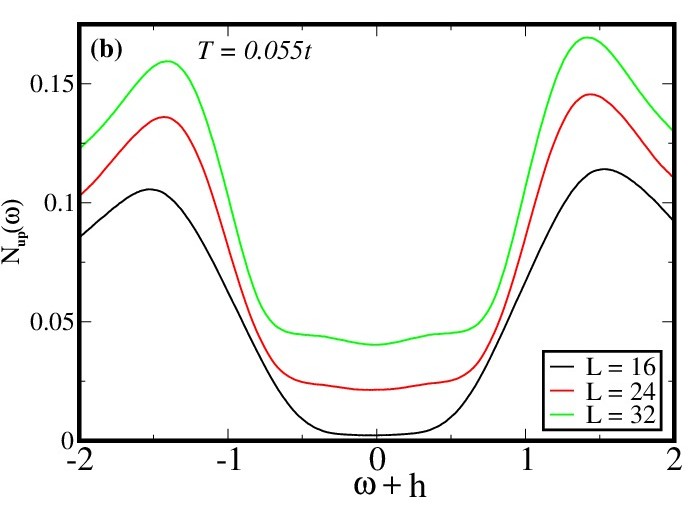}
\hspace{-0.27cm}
\includegraphics[width=5.5cm,height=5.0cm]{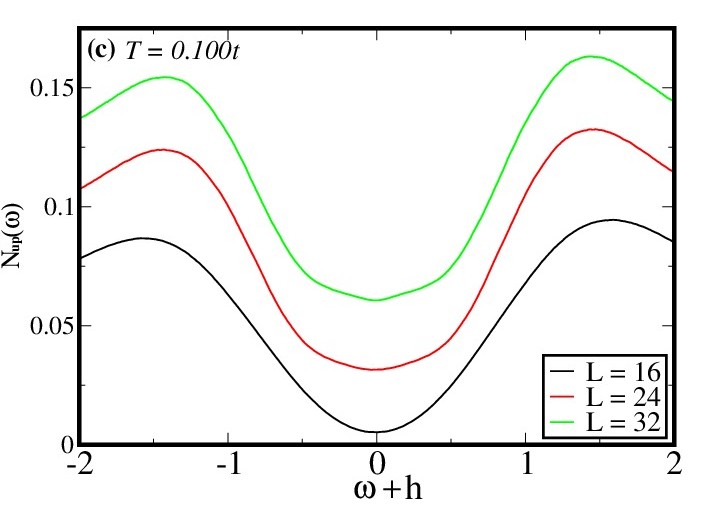}
}
\caption{Size dependence of the DOS at a polarization $P=0.01$
and three temperatures.
The curves are $y$ shifted by equal amounts for clarity.
We observe that (i)~$L=24$ and $L=32$ results are quite similar
(and different from $L=16$), and (ii)~the DOS at low temperature,
panels (a) and (b), have a consistent non monotonic low
energy feature.}
\end{figure*}

The $P \sim 0.075$ trajectory transits from a high $T$ BP phase
to the PPFL. 
The slope of the $h(T)$ curve is much smaller than for $P =0.01$
due to the weaker suppression of Pauli susceptibility.
The $P \sim 0.30$ system passes from a low $T$ LO
phase to a PPFL, while the $P \sim 0.375$ path is always in a
unordered phase but displays a high $T$ pseudogap due
to thermal fluctuations.  For these two cases, where the
ground state does not have
Pauli susceptibility suppression, the $h(T)$ curves generally
have a weak positive slope. In this window the `mysterious'
effect is the presence of a high $T$ pseudogap above a gapless
low temperature state. Addressing this requires a look at the
pairing field distributions.

\subsection{Pairing field distribution}

Fig.4 shows the distribution of the pairing field
for the different fixed $P$ cross sections shown in Fig.2.
Fig.4.(a) shows the pairing field distributions for 
DOS shown in Fig.2.(a), and so on for the other panels.

The anomalous low energy features in Fig.2(a), we believe,
arise from the relatively large number of {\it small $\Delta_i$}
sites in a typical configuration in this part of parameter space.
This arises due to the proximity to the low $\Delta$ FFLO
state. While the mean $\Delta$ is $\sim \Delta_0$ (the $T=0$
mean field value), at $T \sim 0.04t$ in Fig.4(a), 
if the system were fully in the LO regime, Fig.4(c), the
mean $\Delta$ would be $\sim 0.5 \Delta_0$, see the plot
for $T = 0.05t$. We think that proximity to the LO
phase boundary makes the system have a small fraction of
small $\Delta$ sites whose magnitude is $\sim 0.5$ the
typical value. As a result the DOS generates weight
at an energy $\omega_{max}$ that is roughly $0.5$ the
gap edge value. Since the $P=0.01$ trajectory is the
only one that borders the BP-LO boundary the effect 
is not visible at other $P$.

Fig.4.(b) shows a monotonic increase in the width of $P(\vert 
\Delta_i \vert)$ with increasing $T$,
and the increasing `disorder' in the pairing field 
steadily weakens the  PG in the DOS and increases the
low
energy spectral weight. In Fig.4.(c) the lowest $T$ shows the `multimode'
distribution of the amplitude modulated LO state.
At this, and the next
higher $T$, the LO modulations lead to a `pseudogapped' phase but with 
the loss of order at the first order transition, $T \sim 0.005t$, the
mean amplitude collapses. A gapless state emerges, and survives to some
larger temperature, where the thermal fluctuations have created enough
large $\Delta_i$ sites to generate a weak pseudogap. The behavior
in 4.(d) is essentially similar to 4.(c), with the LO part removed:
in this case one just has a gapless to weak PG crossover.  

\subsection{Earlier work}

Much of the earlier work on pseudogaps in imbalanced
superfluids is in the continuum context.
A {\it thermodynamic re-entrance} 
has been pointed out in the continuum unitary gas  
\cite{levin_prl2006,levin2006} at low $P$. 
The authors there had carried out an extensive investigation
in terms of interaction strength $(1/k_Fa)$, polarization, and
temperature and observed that near  $1/k_Fa = 0$ the low $P$
superfluid state has an upper and lower bound in temperature
\cite{levin_prl2006, levin2006, duan}.
At $T = 0$ there is no stable superfluid, while an intermediate 
SF phase is realized at $T \neq 0$.
The superfluid is gapped (by construction in this theory) 
the higher and lower $T$ 
`normal' states are pseudogapped.
The present theory allows for pairing field fluctuations,
consequently we find that within the SF phase the spectrum 
is not always gapped out completely and there are finite low 
energy weight even at the lowest accessible temperature.
Despite the different model, a quite different approximation,
and a thermodynamic stability (rather than spectral
change) argument offered by the authors, we think the
similarity with our result 
is not coincidental. The changing spectral character at the small
polarization regime bears closer experimental search.

The spectrum in imbalanced superconductors can be probed by
tunneling but we are not aware of such studies. Spectral studies
are
more visible in cold atomic superfluids, although the results
are complicated by the effect of a trap. The experiments measure 
a `radio frequency' (RF) current
\cite{shin_prl2007,chin2004,ketterle_prl2008},  rather than the 
density of states.  However, the RF response $I_{\sigma}(\omega)$
is a weighted sum \cite{veillette,levin2008} 
of the same spectral function $A_{\sigma}({\bf k}, \omega)
= -(1/\pi)ImG_{\sigma}({\bf k}, \omega)$ that defines our 
density of states $N_{\sigma}(\omega)$.

Experiments on the unitary Fermi gas indicate 
that the $T_c$ falls from $\sim 0.2T_F$ at $P=0$ 
to zero at $P_c \sim 0.75$. The `pair formation'
scale $T_{pair}$, however, seems to be  ${\cal{O}}(T_F)$ 
{\it even when $P$ crosses $P_c$ and $T_c$ falls to zero} 
\cite{ketterle2007}.
These experiments suggest that a PG is obtained over a
wide temperature window all the way from $P=0$ to
the highest $P$. Our results agree with this, 
except at the high $P$ end where we observe a gapless low
temperature state.
This difference arises from our
neglect of quantum fluctuations in the $\Delta_i(\tau)$,
which can generate a non trivial Fermi liquid ground state
\cite{veillette,levin2008}, preserving pairing but suppressing
condensation.

\subsection{Finite size effect}

The results discussed in this paper correspond to a lattice size 
of $L = 32$. The system size is reasonable in comparison to the 
existing literature in which the thermal physics of spin imbalanced
system are being explored \cite{scaletter}. In order to verify the 
robustness of the spectral features, we have further computed 
our results on various other system sizes. 
Fig.(5) shows the spin resolved DOS calculated at three 
different lattice sizes of $L = 16, 24$ and $32$, for a fixed 
polarization, highlighting three different temperature regimes.
We observed that the anomalous behavior of the spin 
resolved DOS is fairly robust and persists even at $L = 24$,
and is not an artifact of finite size effect.

\section{Conclusion}

We have established the thermal phase diagram of 
population imbalanced lattice fermions near the BCS-BEC 
crossover and observed 
that although the thermodynamic phases exhibit expected
behavior with temperature the spectral features
are very counter-intuitive.
The low temperature superfluid  at a fixed small polarization
has a pseudogap with a distinct maximum in the density of states
in the subgap region, crossing over to a more conventional
featureless pseudogap with increasing temperature.
The unusual low (but finite) temperature result arises due
to LO like fluctuations in the BP phase near the BP-LO boundary.
At large polarization, in the LO phase, the system undergoes
a pseudogap to gapless to pseudogap crossover with increasing
temperature, At even larger 
polarization, where the ground state is gapless and
has no pairing and long range order, heating the system
leads to pseudogap formation above a certain temperature.
All these effects are a consequence of thermal fluctuations
beyond the mean field scheme usually used to analyze these
models.

We acknowledge use of the HPC clusters at HRI and thank
Nyayabanta Swain for comments. 


\end{document}